\documentclass[conference]{IEEEtran}
\IEEEoverridecommandlockouts
\usepackage[utf8]{inputenc}
\usepackage{amsmath,amssymb,amsfonts}
\usepackage{cite}

\usepackage{algorithmic}
\usepackage{stfloats}
\usepackage{graphicx}
\usepackage{textcomp}
\usepackage{xcolor}
\usepackage{pifont}
\usepackage{amsthm}

\usepackage{mathrsfs}
\usepackage{booktabs}
\usepackage{cases}
\usepackage{comment}
\usepackage{empheq} 
\usepackage{caption}
\usepackage{subcaption}
\usepackage{fancyhdr}

\allowdisplaybreaks

\usepackage{url}  
\usepackage{graphicx}  
\usepackage[ruled,vlined,linesnumbered]{algorithm2e}
\usepackage{setspace}
\usepackage{floatpag} 
\usepackage{float}
\floatpagestyle{empty}

\def\BibTeX{{\rm B\kern-.05em{\sc i\kern-.025em b}\kern-.08em
    T\kern-.1667em\lower.7ex\hbox{E}\kern-.125emX}}
\begin{document}
\captionsetup[figure]{labelfont={rm},labelformat={default},labelsep=period,name={Fig.}}
\title{Sum-Rate Maximization in Distributed Intelligent Reflecting Surfaces-Aided mmWave Communications}
%
%
%

\author{Yue Xiu\textsuperscript{1}, Wei Sun\textsuperscript{2}, Jiao Wu\textsuperscript{3}, Guan Gui\textsuperscript{4}, Ning Wei\textsuperscript{1}, Zhongpei Zhang\textsuperscript{1}\\
\IEEEauthorblockA{\textsuperscript{1} National Key Laboratory of Science and Technology on Communications, UESTC, Chengdu, China\\
\textsuperscript{2}School of Computer Science and Engineering, Northeastern University, Shenyang, China\\
\textsuperscript{3}College of Electrical and Computer Engineering,
Seoul National University,
Seoul, South Korea\\
\textsuperscript{4} College of Telecommunications and Information Engineering, NJUPT, Nanjing, China\\
[-10pt] 
}
\thanks{This work was supported in part by the Guangdong province Key Project of science and Technology (2018B010115001), the National Natural Science Foundation of China (NSFC) under Grant 91938202 and 61871070, and the Defense Industrial Technology Development Program (JCKY2016204A603). The corresponding author is Ning Wei.}}
\maketitle

\thispagestyle{fancy}
\pagestyle{fancy}
\lhead{This paper appears in 2021 IEEE Wireless Communications and Networking Conference (WCNC 2021). Please feel free to contact us for questions or remarks.}
\cfoot{\thepage}
\renewcommand{\headrulewidth}{0.4pt}
\renewcommand{\footrulewidth}{0pt}
\begin{abstract}
In this paper, we focus on the sum-rate optimization in a multi-user millimeter-wave (mmWave) system with distributed intelligent reflecting surfaces (D-IRSs), where a base station (BS) communicates with users via multiple IRSs. The BS transmit beamforming, IRS switch vector, and phase shifts of the IRS are jointly optimized to maximize the sum-rate under minimum user rate, unit-modulus, and transmit power constraints. To solve the resulting non-convex optimization problem, we develop an  efficient  alternating optimization (AO) algorithm. Specifically, the non-convex problem is converted into three subproblems, which are solved alternatively. The solution to transmit beamforming at the BS and the phase shifts at the IRS are derived by using the successive convex approximation (SCA)-based algorithm, and a greedy algorithm is proposed to design the IRS switch vector. The complexity of the proposed AO algorithm is analyzed theoretically. Numerical results show that the D-IRSs-aided scheme can significantly improve 
the sum-rate and energy efficiency performance.
\end{abstract}


%
\IEEEpeerreviewmaketitle

\section{Introduction}
Millimeter-wave (mmWave) is widely acknowledged as a promising technology for the fifth-generation (5G) communications, which can achieve ultra-high data-rate \cite{di2019smart,lu2020robust,pi2011introduction,gui2019new,gui20206g}. 
However, the high path loss and severe blockages in mmWave bands greatly degrades the quality of service (QoS)\cite{dai2018hybrid}.
The intelligent reflecting surface (IRS)~technology has been recently investigated for overcoming the serious path loss and enhancing the data-rate \cite{cao2019intelligent}.
Specifically, IRS is a planar metasurface consisting of a large number of passive reflecting elements, each of which is able to reflect the incident signals with a desired phase shift \cite{wu2019towards}. 
By adaptively altering the propagation of the reflected signal, the IRS is capable of improving the received signal power via constructive signal combination and destructive interference mitigation at the receivers, thereby enhancing the system performance.

Various  studies  over  the  transmit beamforming  and phase shifts of the IRS  design  for  IRS-aided wireless systems have been increasingly made \cite{wu2019towards,wu2019intelligent,guo2019weighted,pan2020multicell}.
In \cite{wu2019towards}, the transmit beamforming at the base station (BS) and the phase shifts at the IRS have been jointly designed to maximize the achievable sum-rate.
In \cite{wu2019intelligent}, an IRS-aided multi-user multiple-input single-output (MISO) system has been studied, where the phase shift matrix and transmit beamforming were jointly optimized by semidefinite relaxation and alternating optimization techniques.
In \cite{guo2019weighted}, the authors studied for a single-user IRS-aided MISO system, and a dual decomposition and price-based method are used to maximize the weighted sum-rate. 
In \cite{pan2020multicell}, a sum-rate maximization problem has been studied, where manifold optimization algorithm was adopted for designing phase shifts of the IRS.
However, these works are mainly oriented to the microwave communications, while mmWave systems with multiple IRSs have still been unexplored. 

These unsolved problems motivate us to investigate the sum-rate optimization problem in a multi-user mmWave system assisted by distributed IRSs (D-IRSs). 
To be specific, the key idea of the proposed alternating optimization (AO) scheme is to jointly optimize the phase shiftsat the IRS, the switch vector at the IRS and the transmit beamforming at the BS, subject to the constraints on the transmit power and user rate. 
Due to the non-convexity of the optimization problem, we propose a novel algorithm to maximize the sum-rate for a D-IRSs aided multi-user MISO system, referred to as the joint beamforming, switch and phase shifts optimization algorithm.
In the proposed algorithm,
the transmit beamforming is firstly derived by the successive convex approximation (SCA) algorithm, 
then which is also used to obtain the phase shift matrix at the IRS. 
For the design of the IRS switch vector, a greedy algorithm is exploited.
Finally, we demonstrate from numerical results that the proposed algorithm 
can achieve high sum-rate while improving the energy efficiency.


\section{System Model}
\begin{figure}[!t]
\centering
\includegraphics[height=2in,width=3in]{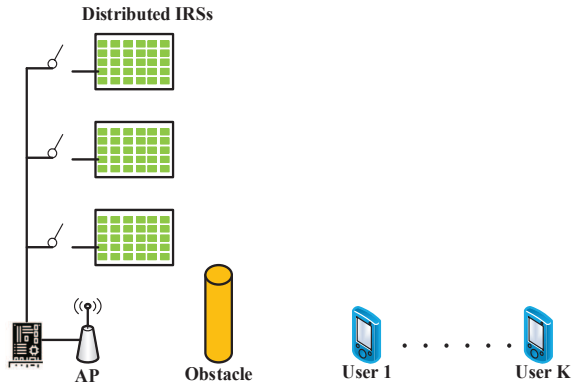}
\caption{System model for mmWave communication system with D-IRSs.\vspace{-8pt}}
\label{fig:1}
\end{figure}
Consider the system model in Fig.~\ref{fig:1}, where a $N_{t}$-antenna BS transmits signals to $K$ single-antenna users. 
This communication is aided by $L$ IRS units, where each IRS comprises $N_{r}$ reflecting elements being able to reflect the incident signal independently with an adjustable phase shift. 
In this paper, we assume that the direct link between the BS and each user is blocked by obstacles. 
Also, $L$ IRSs 
are assumed to be deployed on high buildings around the users.
Thus, the BS-IRS channel is dominated by the line of sight (LoS) path, which can be expressed as
\begin{align}
\boldsymbol{G}_{l}=\sqrt{\frac{1}{\beta_{l}}}\alpha_{l}\boldsymbol{a}\boldsymbol{b}^{H},\label{6-1}
\end{align}
where $\beta_{l}$ is the large-scale fading coefficient for the $l$th path, and $\alpha_{l}\sim\mathcal{CN}(0,1)$ is the small-scale fading coefficient. $\boldsymbol{b}\in\mathbb{C}^{N_{t}\times 1}$ and  $\boldsymbol{a}\in\mathbb{C}^{N_{r}\times 1}$ are the array response {}{vector} at the BS and IRS, respectively. 
The channel from {}{the} $l$th IRS to the $k$th user is expressed as
\begin{align}
\boldsymbol{h}_{kl}=\sqrt{\frac{1}{(\beta_{kl}L_{kl})}}\sum\nolimits_{L=0}^{L_{kl}-1}\alpha_{kl}\boldsymbol{a}_{kl},\label{6-2}
\end{align}
where $\beta_{kl}$ and $\alpha_{kl}\sim\mathcal{CN}(0,1)$ are similarly defined as that in (\ref{6-1}). $L_{kl}$ is the number of paths from the $l$th IRS to the $k$th user, $\boldsymbol{a}_{kl}$ is the transmit array steering vector at the IRS. 
In this work, the channel state information (CSI) is assumed to be perfectly known at {}{BS}. 

In the {}{D-IRSs}-aided mmWave system, 
The received signal at {}{the} $k$th user can be written as
\begin{align}
y_{k}=&\sum\nolimits_{l=1}^{L}x_{l}\boldsymbol{h}^{H}_{kl}\boldsymbol{\Theta}_{l}\boldsymbol{G}_{l}\boldsymbol{w}_{k}s_{k}+ \sum\nolimits_{i\neq k}^{K}\sum\nolimits_{l=1}^{L}x_{l}\boldsymbol{h}^{H}_{kl}\boldsymbol{\Theta}_{l}\boldsymbol{G}_{l}\boldsymbol{w}_{i}\nonumber\\
&s_{i}+n_{k},\label{6-3}
\end{align}
where $\boldsymbol{s}=[s_{1},\ldots,s_{K}]\in\mathbb{C}^{K\times 1}$ is the transmit signal, satisfying $\mathbb{E}\{s_{k}s_{k}^{H}\}=1$, $\mathbb{E}\{s_{i}s_{j}^{H}\}=0, i\neq j$, and $n_{k}\sim\mathcal{CN}(0,\sigma^{2}_{k})$ is the noise vector. $\boldsymbol{\Theta}_{l}=\mathrm{diag}\{e^{j\theta_{l1}},\cdots,e^{j\theta_{lN_{r}}}\}$ {}{denotes} the reflecting matrix of the $l$th IRS. 
$\boldsymbol{x}=[x_{1},\ldots,x_{L}]^{T}$ is defined as the switch vector with $x_{l}\in\{0,1\}$. 
$x_{l}=1$ means that the $l$th IRS is active, while $x_{l}=0$ represents that the $l$th IRS deos not work and consume any power. 

Then, the signal to interference plus noise ratio (SINR) at the user $k$ can be {defined} as
\begin{align}
\mathrm{SINR}_{k}=\frac{|\sum\nolimits_{l=1}^{L}x_{l}\boldsymbol{h}^{H}_{kl}\boldsymbol{\Theta}_{l}\boldsymbol{G}_{l}\boldsymbol{w}_{k}|^{2}}{\sum\nolimits_{i\neq k}^{K}|\sum\nolimits_{l=1}^{L}x_{l}\boldsymbol{h}^{H}_{kl}\boldsymbol{\Theta}_{l}\boldsymbol{G}_{l}\boldsymbol{w}_{i}|^{2}+\sigma_{k}^{2}}.\label{6-4}
\end{align}

In this paper, we aim to maximize the sum-rate by jointly designing the phase shift matrix and the switch vector at the IRS, and the transmit beamforming at the BS, under the constraints of user rate and the transmit power. 
Therefore, the optimization problem is formulated as
\begin{subequations}
\begin{align}
(\text{P1}): \max_{\{\boldsymbol{w}_{k}\},\{\boldsymbol{\Theta}_{l}\},\boldsymbol{x}}~&\sum\nolimits_{k=1}^{K}R_{k},\label{6-5a}\\
\mbox{s.t.}~
&R_{k}\geq \gamma_{k},\label{6-5b}\\
&\sum\nolimits_{k=1}^{K}\mathrm{Tr}\left(\boldsymbol{w}_{k}\boldsymbol{w}_{k}^{H}\right)\leq P,\label{6-5c}&\\
&|\theta_{l,j}|=1,~\forall l\in\mathcal{L},j=1,\cdots,N_{r}.&\label{6-5d}\\
&x_{l}\in\{0,1\}, \forall l\in\mathcal{L},&\label{6-5e}
\end{align}\label{6-5}%
\end{subequations}
where $R_{k}=\log_{2}(1+\mathrm{SINR}_{k})$ is defined as the achievable rate of the user $k$, $P$ is the maximum transmit power. 
It is easily find that the problem (P1) is highly non-convex due to the non-convexity of the objective function and constraints, which is challenging to solve.


\section{Sum-Rate Maximization Via the Alternating Optimization Algorithm}
In this section, we propose a new scheme to handle the non-convex problem (P1). 
Firstly, we optimize the transmit beamforming vector $\boldsymbol{w_{k}}$ and the phase shift matrix $\boldsymbol{\Theta}_{l}$ by using the SCA algorithm. 
Then, the IRS switch vector $\boldsymbol{x}$ is optimized via a greedy algorithm. 
\subsection{Transmit Beamforming Design}
Given the phase shift matrix $\boldsymbol{\Theta}_{l}$ and the IRS switch vector $\boldsymbol{x}$, (P1) can be rewritten as
\begin{subequations}
\begin{align}
(\text{P2}): \max_{\{\boldsymbol{w}_{k}\}}~&\sum\nolimits_{k=1}^{K}R_{k},\label{6-6a}\\
\mbox{s.t.}~
&R_{k}\geq \gamma,\label{6-6b}\\
&\sum\nolimits_{k=1}^{K}\mathrm{Tr}\left(\boldsymbol{w}_{k}\boldsymbol{w}_{k}^{H}\right)\leq P.\label{6-6c}
\end{align}\label{6-6}%
\end{subequations}

In order to make the problem (P2) more tractable, we introduce two new variables $\boldsymbol{W}_{k}=\boldsymbol{w}_{k}\boldsymbol{w}_{k}^{H}$, and $\boldsymbol{a}^{H}_{k}=\sum\nolimits_{l=1}^{L}x_{l}\boldsymbol{h}_{kl}^{H}\boldsymbol{\Theta}_{l}\boldsymbol{G}_{l}$.
Then, we have $\|\sum\nolimits_{l=1}^{L}x_{l}\boldsymbol{h}_{kl}^{H}\boldsymbol{\Theta}_{l}\boldsymbol{G}_{l}\boldsymbol{w}_{k}\|^{2}=\mathrm{Tr}(\boldsymbol{W}_{k}\boldsymbol{A}_{k})$, and $\|\sum\nolimits_{l=1}^{L}x_{l}\boldsymbol{h}_{kl}^{H}\boldsymbol{\Theta}_{l}\boldsymbol{G}_{l}\boldsymbol{w}_{i}\|^{2}=\mathrm{Tr}(\boldsymbol{W}_{i}\boldsymbol{A}_{k})$, with $\boldsymbol{A}_{k}=\boldsymbol{a}_{k}\boldsymbol{a}_{k}^{H}$ and $\mathrm{rank}(\boldsymbol{W}_{k})=1$.
Here, we exploit the semidefinite relaxation (SDR) to drop the rank-one constraint $\mathrm{rank}(\boldsymbol{W}_{k})=1$.
Therefore, (P2) can be expressed as
\begin{subequations}
\begin{align}
(\text{P3}): \max_{\{\boldsymbol{W}_{k}\}}~&\sum\nolimits_{k=1}^{K}\log_{2}\left(1+\frac{\mathrm{Tr}(\boldsymbol{W}_{k}\boldsymbol{A}_{k})}{\sum\nolimits_{i\neq k}^{K}\mathrm{Tr}(\boldsymbol{W}_{i}\boldsymbol{A}_{k})+\sigma^{2}_{k}}\right),\label{6-7a}\\
\mbox{s.t.}~
&\log_{2}\left(1+\frac{\mathrm{Tr}(\boldsymbol{W}_{k}\boldsymbol{A}_{k})}{\sum\nolimits_{i\neq k}^{K}\mathrm{Tr}(\boldsymbol{W}_{i}\boldsymbol{A}_{k})+\sigma^{2}_{k}}\right)\geq \gamma,\label{6-7b}\\
&\sum\nolimits_{k=1}^{K}\mathrm{Tr}\left(\boldsymbol{W}_{k}\right)\leq P,&\label{6-7c}\\
&\boldsymbol{W}_{k}\succeq\boldsymbol{0}.\label{6-7d}
\end{align}\label{6-7}%
\end{subequations}

However, the problem (P3) is still non-convex due to the non-convex objective function (\ref{6-7a}). 
In order to transform (P3) into a convex problem, we introduce new variables
\begin{eqnarray}
e^{p_{i}}=\sum\nolimits_{k=1}^{K}\mathrm{Tr}(\boldsymbol{W}_{k}\boldsymbol{A}_{i})+\sigma_{k}^{2}\label{6-8}\\
e^{q_{i}}=\sum\nolimits_{k\neq i}^{K}\mathrm{Tr}(\boldsymbol{W}_{k}\boldsymbol{A}_{i})+\sigma_{k}^{2}.\label{6-9}
\end{eqnarray}
Then, we have
\begin{subequations}
\begin{align}
(\text{P4}): \max_{\{\boldsymbol{W}_{k}\},\{p_{i}\},\{q_{i}\}}~&\sum\nolimits_{i=1}^{K}\log_{2}\left(e^{p_{i}-q_{i}}\right),\label{6-10a}\\
\mbox{s.t.}~
&(p_{i}-q_{i})\log_{2}(e)\geq\gamma,\label{6-10b}\\
&\sum\nolimits_{k=1}^{K}\mathrm{Tr}(\boldsymbol{W}_{k}\boldsymbol{A}_{i})+\sigma_{k}^{2}\geq e^{p_{i}},&\label{6-10c}\\
&\sum\nolimits_{k\neq i}^{K}\mathrm{Tr}(\boldsymbol{W}_{k}\boldsymbol{A}_{i})+\sigma_{k}^{2}\leq e^{q_{i}},&\label{6-10d}\\
&\sum\nolimits_{k=1}^{K}\mathrm{Tr}\left(\boldsymbol{W}_{k}\right)\leq P,\boldsymbol{W}_{k}\succeq\boldsymbol{0},&\label{6-10e}\\
&\mathrm{Tr}({\boldsymbol{W}_{k}\boldsymbol{A}_{i}})\geq 0.&\label{6-10f}
\end{align}\label{6-10}%
\end{subequations}
We observe that
\begin{align}
\sum\nolimits_{i=1}^{K}\log_{2}\left(e^{p_{i}-q_{i}}\right)=\sum\nolimits_{i=1}^{K}(p_{i}-q_{i})\log_{2}(e),\label{6-11}
\end{align}
thus, the objective function (\ref{6-10a}) is  convex.

By replacing (\ref{6-10c}) and (\ref{6-10d}) with (\ref{6-8}) and (\ref{6-9}) with, we see that the inequalities (\ref{6-10c}) and (\ref{6-10d}) hold with equalities when the solution is optimal. 
The main reason is that the objective function in the problem (P4) is monotonous.
In this case, we aim to maximize $e^{p_{i}}$ while minimizing $e^{q_{i}}$, rather than directly maximizing the objective function (\ref{6-10a}).


Next, we use the successive convex approximation (SCA) algorithm to solve the problem (P4). 
The first-order Taylor expansion of $e^{q_{i}}$ at the point $\bar{q}_{i}$ is given by
\begin{eqnarray}
e^{\bar{q}_{i}}+e^{\bar{q}_{i}}(q_{i}-\bar{q}_{i}),\label{6-12}
\end{eqnarray}
where $\bar{q}_{i}$ is feasible to the problem (P4). 
And the constraint in (\ref{6-10d}) can be rewritten as
\begin{eqnarray}
\sum\nolimits_{k\neq i}^{K}\mathrm{Tr}(\boldsymbol{W}_{k}\boldsymbol{A}_{i})+\sigma_{k}^{2}\leq e^{\bar{q}_{i}}+e^{\bar{q}_{i}}(q_{i}-\bar{q}_{i}).\label{6-13}
\end{eqnarray}

It is observed that (\ref{6-13}) is convex since that (\ref{6-12}) is linear and convex.
Then, by replacing (\ref{6-10d}) with (\ref{6-13}), we have
\begin{subequations}
\begin{align}
(\text{P5}):
\max_{\boldsymbol{W},\{p_{i}\},\{q_{i}\}}~&\sum\nolimits_{i=1}^{K}\log_{2}\left(e^{p_{i}-q_{i}}\right),\label{6-14a}\\
\mbox{s.t.}~
&\sum\nolimits_{k\neq i}^{K}\mathrm{Tr}(\boldsymbol{W}_{k}\boldsymbol{A}_{i})+\sigma_{k}^{2}\notag\\
&\leq e^{\bar{q}_{i}}+e^{\bar{q}_{i}}(q_{i}-\bar{q}_{i}),&\label{6-14b}\\
&\text{(\ref{6-10b})},\text{(\ref{6-10c})},\text{(\ref{6-10e})},\text{(\ref{6-10f})}.&\label{6-14c}
\end{align}\label{6-14}%
\end{subequations}

Note that (P5) is a convex optimization problem that can be solved by using the convex optimization toolbox, e.g. CVX\cite{boyd2004convex}. 

The SCA-based algorithm for solving (P5) is summarized in \textbf{Algorithm}~\ref{algo-1}\footnote{In practical mmWave systems, the transmitter is usually equipped with the hybrid beamforming structure. After obtaining $\boldsymbol{W}=[\boldsymbol{w}_{1},\cdots,\boldsymbol{w}_{K}]$ from \textbf{Algorithm}~\ref{algo-1}, therefore, we use OMP algorithm to design the hybrid beamforming \cite{tropp2007signal}.}.

\begin{algorithm}
\caption{Proposed SCA-based Algorithm for Solving the Problem (P2).} 
\label{algo-1}
\hspace*{0.02in}{\bf Initialization:} $t=0$,  
given $\boldsymbol{w}^{0}$ that is satisfied conditions, calculate $q^{0}_{i}$ based on (\ref{6-9}) and let $\bar{q}_{i}^{1}=q^{0}_{i}$.\\
\hspace*{0.02in}{\bf Repeat}\\
Solve the problem in (\ref{6-14}) to obtain the optimal solution $\{\boldsymbol{W}_{k}^{t}\}$ and $\{q_{i}^{t}\}$.\\
Update $\bar{q}_{i}^{t+1}=q_{i}^{t}$.\\
Set $t=t+1$.\\
\hspace*{0.02in}{\bf Until}
The stopping criterion is met.\\
\hspace*{0.02in}{\bf Output:} 
Obtain $\boldsymbol{w}_{k}^{*}$ by decomposition of $\boldsymbol{W}_{k}^{*}$ when the $\mathrm{rank}(\boldsymbol{W}_{k}^{*})=1$; otherwise the Gaussian Randomization method {is} utilized to obtain a rank-one approximation.\\
\end{algorithm}

\subsection{Phase Shifts Optimization}
Given the beamforming vector $\{\boldsymbol{w}_{k}\}$ and the IRS shift switch vector $\boldsymbol{x}$, the problem in (\ref{6-6}) can be expressed as 
\begin{subequations}
\begin{align}
(\text{P6}):
\max_{\boldsymbol{\lambda},\{\boldsymbol{\Theta}_{l}\}}~&\sum\nolimits_{k=1}^{K}\log_{2}(1+\lambda_{k})  \label{6-16a}\\
\mbox{s.t.}~
&\lambda_{k}\leq \frac{|\sum\nolimits_{l=1}^{L}\boldsymbol{h}^{H}_{kl}x_{l}\boldsymbol{\Theta}_{l}\boldsymbol{G}_{l}\boldsymbol{w}_{k}|^{2}}{\sum\nolimits_{i\neq k}^{K}|\sum\nolimits_{l=1}^{L}\boldsymbol{h}^{H}_{kl}x_{l}\boldsymbol{\Theta}_{l}\boldsymbol{G}_{l}\boldsymbol{w}_{i}|^{2}+\sigma_{k}^{2}},\label{6-16b}\\
&\lambda_{k}\geq 2^{\gamma}-1,\label{6-16c}\\
&|\theta_{l,j}|=1,\label{6-16d}
\end{align}\label{6-16}%
\end{subequations}
where $\boldsymbol{\lambda}=[\lambda_{1},\cdots,\lambda_{K}]^{T}$, $\lambda_{i}$ is a slack variable, ensuring that the constraint (\ref{6-16b}) always
holds with equality for the optimal solution. 

Let $u_{ln}=e^{j\theta_{ln}}$, $\boldsymbol{u}_{l}=[u_{l1},\cdots,u_{lN_{r}}]^{T}$, and $\boldsymbol{u}=[u_{11},\cdots,u_{1N_{r}},\cdots,u_{LN_{r}},\cdots,u_{LN_{r}}]^{T}$. 
Since $x_{l}\boldsymbol{h}_{kl}^{H}\boldsymbol{\Theta}_{l}\boldsymbol{G}_{l}\boldsymbol{w}_{i}=\boldsymbol{v}_{kli}^{H}\boldsymbol{u}_{l}$ with $\boldsymbol{v}_{kli}=x_{l}(\mathrm{diag}(\boldsymbol{h}_{kl}^{H})\boldsymbol{G}_{l}\boldsymbol{w}_{i})^{*}$, the constraint (\ref{6-16b}) can be rewritten as
\begin{align}
\lambda_{k}\leq\frac{|\boldsymbol{v}_{kk}^{H}\boldsymbol{u}|^{2}}{\sum\nolimits_{i\neq k}^{K}|\boldsymbol{v}_{ki}^{H}\boldsymbol{u}|^{2}+\sigma^{2}},\label{6-17}
\end{align}
where $\boldsymbol{v}_{ki}=[\boldsymbol{v}_{ki1},\cdots,\boldsymbol{v}_{kiL}]$. 
Then, (\ref{6-17}) can be transformed into
\begin{align}
\lambda_{k}(\sum\nolimits_{i\neq k}^{K}|\boldsymbol{v}_{ki}^{H}\boldsymbol{u}|^{2}+\sigma^{2})-|\boldsymbol{v}_{kk}^{H}\boldsymbol{u}|^{2}\leq 0.\label{6-17m}
\end{align}

Therefore, the problem (P6) can be rewritten as
\begin{subequations}
\begin{align}
(\text{P7}):
\max_{\boldsymbol{u},\boldsymbol{\lambda}}~&\sum\nolimits_{k=1}^{K}\log_{2}(1+\lambda_{k}),\label{6-18a}\\
\mbox{s.t.}~
&|u_{ln}|=1,\label{6-18b}\\
&\text{(\ref{6-16c})},\text{(\ref{6-17m})}.\label{6-18c}
\end{align}\label{6-18}%
\end{subequations}

By introducing the penalty factor, (\ref{6-18}) can be reformulated as
\begin{subequations}
\begin{align}
(\text{P8}):
\max_{\boldsymbol{u},\boldsymbol{\lambda}}~&\sum\nolimits_{k=1}^{K}\log_{2}(1+\lambda_{k}) \notag\\
&+\mu\sum\nolimits_{l=1}^{L}\sum\nolimits_{n=1}^{N_{r}}(|u_{ln}|^{2}-1),\label{6-19a}\\
\mbox{s.t.}~
&|u_{ln}|\leq 1,\label{6-19b}\\
&\text{(\ref{6-16c})},\text{(\ref{6-17m})},\label{6-19c}
\end{align}\label{6-19}%
\end{subequations}
where $\mu$ is a large positive constant. 
In order to solve the non-convex parts in (\ref{6-17m}) and (\ref{6-19a}), the SCA algorithm is used.
The first-order Taylor series of (\ref{6-19a}) and (\ref{6-17m}) can be respectively expressed as
\begin{align}
\sum\nolimits_{k=1}^{K}\log_{2}(1+\lambda_{k})+\mu\sum\nolimits_{l=1}^{L}\sum\nolimits_{n=1}^{N_{r}}u_{ln}^{t}(u_{ln}-u_{ln}^{t}), \label{6-20}
\end{align}
and
\begin{align}
&\lambda_{k}(\sum\nolimits_{i\neq k}^{K}|\boldsymbol{v}_{ki}^{H}\boldsymbol{u}|^{2}+\sigma^{2})-|\boldsymbol{v}_{kk}^{H}\boldsymbol{u}^{t}|^{2}-2\mathrm{Re}[(\boldsymbol{u}^{H})^{t}\boldsymbol{v}_{kk}\boldsymbol{v}_{kk}^{H}\nonumber\\
&(\boldsymbol{u}-\boldsymbol{u}^{t})]\leq 0.\label{6-20m}
\end{align}

With the above approximations at hand, the non-convex problem (P8) can be rewritten as
\begin{subequations}
\begin{align}
(\text{P9}):
\max_{\boldsymbol{u},\boldsymbol{\lambda}}~&\sum\nolimits_{k=1}^{K}\log_{2}(1+\lambda_{k})+\mu\sum\nolimits_{l=1}^{L}\sum\nolimits_{n=1}^{N_{r}}u_{ln}^{t}(u_{ln}-u_{ln}^{t}),\label{6-24a}\\
\mbox{s.t.}~
&\text{(\ref{6-19b})},\text{(\ref{6-16c})},\text{(\ref{6-20m})}.\label{6-24c}
\end{align}\label{6-24}%
\end{subequations}
The procedure for solving the problem (P9) is similar to $\textbf{Algorithm}~\ref{algo-1}$.

\subsection{IRS Switch Optimization}
Given the phase shifts matrix $\{\boldsymbol{\Theta}_{l}\}$ and the beamforming vector $\{\boldsymbol{w}_{k}\}$, the problem (\ref{6-6}) is a nonlinear integer optimization problem with respect to $\boldsymbol{x}$. 
Since the nonlinear integer optimization problem is NP-hard in general, it is difficult to obtain the globally optimal solution. 
Thus, we propose a greedy method here, which is explained in \textbf{Algorithm}~\ref{algo-2}. 
\begin{algorithm}
\caption{Proposed Greedy Algorithm for $\boldsymbol{x}$.} 
\label{algo-2}
\hspace*{0.02in}{\bf Initialization} $\mathcal{S}=\{1,\cdots,L\}$ and $x_{l}=1$, $\forall l\in\mathcal{S}$.\\
Calculate the objective function value in (\ref{6-6a}), which is denoted by $V_{0}$.\\
\hspace*{0.02in}{\bf If} $\mathcal{S}\neq \emptyset$.\\
\hspace*{0.02in}{\bf For} $l\in\mathcal{S}$.\\
Turn off the $l$th IRS, i.e., $x_{l}=0$, $x_{m}=1$, $x_{n}=0$, $\forall m\in\mathcal{S}\setminus\{l\}$, $n\in\mathcal{L}\setminus \mathcal{S}$.\\
{}{When $R_{k}\geq \gamma$, $\forall~k$, calculate (\ref{6-6a}), which is denoted by $V_{l}$.}\\
{}{When $R_{k}< \gamma$, $\exists~k$, set $V_{l}=0$.}\\
\hspace*{0.02in}{\bf End}.\\
Calculate $k=\arg\max_{j\in\mathcal{S}\bigcup\{0\}}V_{j}$.\\
\hspace*{0.02in}{\bf If} $k\neq 0$\\
Set $\mathcal{S}=\mathcal{S}\setminus \{k\}$ and $V_{0}=V_{k}$.
\\
\hspace*{0.02in}{\bf Else}\\
Break and jump to Step 16.\\
\hspace*{0.02in}{\bf End}.\\
\hspace*{0.02in}{\bf End}.\\
\hspace*{0.02in}{\bf Output:} 
$x_{l}=1,x_{n}=0,~\forall~l\in\mathcal{S}, n\in\mathcal{L}\setminus \mathcal{S}$.\\
\end{algorithm}

In \textbf{Algorithm}~\ref{algo-2}, $\mathcal{S}$ represents the active IRSs set, and $V^{0}$ is initialized as the objective function.
In the step 5, we deactivate one of the IRSs and obtain a IRS shift switch vector $\boldsymbol{x}$.
If the obtained solution $\boldsymbol{x}$ is feasible, then the value of the objective function can be calculated based on (\ref{6-6a}) in the step 6.
Otherwise, the value of the objective function is set as 0 in the step 7. 
Next, the sum-rates of the newly obtained feasible value and the initial value are compared.
Here, $k\neq 0$ denotes that the sum-rate can be increased by deactivating the IRS, and the highest sum-rate can be achieved by deactivating the IRS $k$ rather than any other IRSs. 
Then, the index $k$ is removed from
the active IRSs set $\mathcal{S}$, and the initial sum-rate valueis updated in the step 11. 
And $k=0$ denotes that deactivating any IRSs will lead to a reduction in the sum-rate. 
In this case, the iteration isterminated since that deactivating any IRSs does not increase the sum-rate. 

Finally, the proposed AO algorithm is summarized in \textbf{Algorithm}~\ref{algo-3}. 
\begin{algorithm}
\caption{Proposed AO Algorithm for Problem (\ref{6-5}).} 
\label{algo-3}
\hspace*{0.02in}{\bf Initialization} $\{\boldsymbol{w}_{k}^{0}\}$, $\boldsymbol{x}^{0}$, and $\{\boldsymbol{\Theta}_{l}^{0}\}$.\\
\hspace*{0.02in}{\bf Repeat}\\
Solve (\ref{6-6}) via \textbf{Algorithm}~\ref{algo-1}.\\
Solve (\ref{6-16}) via SCA-based algorithm.\\
Obtain $\boldsymbol{x}$ via \textbf{Algorithm}~\ref{algo-2}.\\
\hspace*{0.02in}{\bf Until} the objective value (\ref{6-5a}) converges.\\
\hspace*{0.02in}{\bf Output:} 
$\{\boldsymbol{w}_{k}^{0}\}$, $\boldsymbol{x}^{0}$, and $\{\boldsymbol{\Theta}_{l}^{0}\}$.\\
\end{algorithm}
\subsection{Computational Complexity}
According to \cite{grant2009cvx}, the total complexity of the SCA algorithm for solving the problem in (\ref{6-14}) is $\mathcal{O}(S_{1}(2KN_{t}^{2}+2K)^{3.5}\log_{2}(\frac{1}{\epsilon_{1}}))$, where $S_{1}$ is the number of iterations. 
With similar analysis, the total complexity of the SCA algorithm for solving the beamforming optimization problem in (\ref{6-24}) is $\mathcal{O}(S_{2}(2LN_{r}+K)^{3.5}\log_{2}(\frac{1}{\epsilon_{2}}))$, where $\epsilon_{1}$ and $\epsilon_{2}$ are the accuracies of the SCA algorithm for solving problem (\ref{6-14}) and (\ref{6-24}), respectively. $S_{2}$ is the number of iterations. The computation complexity of the proposed greedy algorithm for {}{obtaining IRS switch vector} is $\mathcal{O}(L^{3}N_{r}N_{t})$, {}{where $L$ is the number of variables, and $L^{2}$ is the number of iterations for \textbf{Algorithm}~\ref{algo-2}.}
As a result, the total complexity of the proposed JSTPO algorithm for solving problem (\ref{6-5}) is $\mathcal{O}(TS_{1}(2KN_{t}^{2}+2K)^{3.5}\log_{2}(\frac{1}{\epsilon_{1}})+
TS_{2}(2LN_{r}+K)^{3.5}\log_{2}(\frac{1}{\epsilon_{2}})+TL^{3}N_{r}N_{t})$, where $T$ is the number of iterations for proposed AO algorithm.

\section{Numerical Results}
\begin{figure}[!t]
\centering
\includegraphics[height=2.0in,width=3.0in]{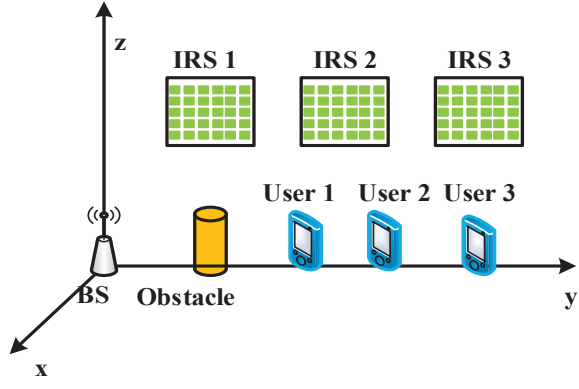}
\caption{Simulation Setup.\vspace{-8pt}}
\label{fig:2}
\end{figure}

In this section, numerical results are provided to demonstrate the effectiveness of the proposed scheme. 
As shown in Fig.~\ref{fig:2}, it is assumed that the BS is located at $(0,0,0)$. 
Three IRSs are located at $(0,30,20)$, $(0,60,20)$, $(0,90,20)$ in meters, respectively. 
While three users are located at $(0,30,0)$, $(0,60,0)$, and $(0,90,0)$ in meters, respectively. 
The large-scale fading $\beta_{l}$ and $\beta_{kl}$ are taken as $\beta_{0}+10c\log_{10}(d)$, where $d$ is the propagation distance of signals, and the path loss exponents of LoS and No-LoS are respectively set as $c=2$ and $c=5$. 
In the simulations, we set $N_{t}=16$, $N_{r}=16$, $L=3$, $\beta_{0}=61.4$ dB, and $\sigma_{k}^{2}=-100$ dBm.  
We compare the proposed scheme with the Single-IRS (S-IRS) scheme and all-active-IRS scheme (D-IRSs without switch). 
In the S-IRS scheme, two scenarios, i.e., the IRS locates at $(0,60,20)$ and $(0,90,20)$ in meters are both considered.
The number of reflecting elements of the IRS is set as the total number of reflecting elements for all IRSs in D-IRSs-aided mmWave system in the S-IRS scheme.
In addition, we evaluate the energy efficiency of the proposed scheme. 
In this paper, the energy efficiency is defined as
\begin{align}
\eta_{EE}=\frac{R_{sum}}{P_{W}+N_{RF}P_{RF}+\sum\nolimits_{l=1}^{L}x_{l}N_{r}P_{IRS}}.\label{24}
\end{align}
where $R_{sum}=\sum\nolimits_{k=1}^{K}R_{k}$ is the sum-rate, $P_{W}=\sum\nolimits_{k=1}^{K}\mathrm{Tr}(\boldsymbol{w}_{k}\boldsymbol{w}_{k}^{H})$ is the transmit power, $P_{RF}$ is the RF chain power, and $N_{r}P_{IRS}$ is the power consumption of each IRS.  
Also, we set $P_{RF}$ and $P_{IRS}$ as $250$ mW and $10$ mW, respectively\cite{huang2019reconfigurable}.
\begin{figure}
\centering
\begin{subfigure}{.3\textwidth}
  \centering
  \includegraphics[scale=0.4]{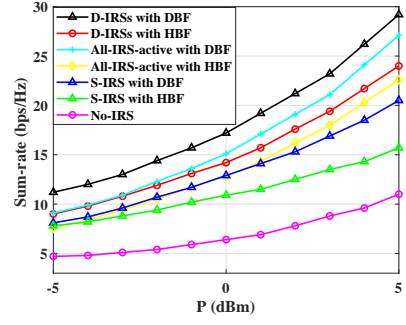}  
  \caption{$N_{r}=16$, $N_{RF}=8$}
  \label{fig:(a1)}
\end{subfigure}
\begin{subfigure}{.3\textwidth}
  \centering
  \includegraphics[scale=0.4]{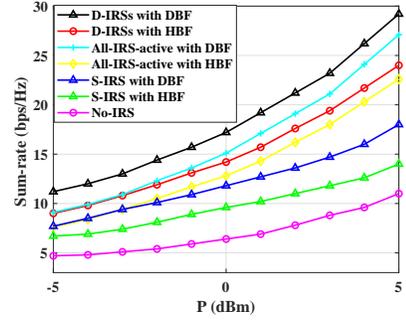}  
  \caption{$N_{r}=16$, $N_{RF}=8$}
  \label{fig:(b1)}
\end{subfigure}
\caption{Sum-rate versus transmit power $P$}
\label{fig:3a}
\end{figure}

\begin{figure}
\centering
\begin{subfigure}{.3\textwidth}
  \centering
  \includegraphics[scale=0.4]{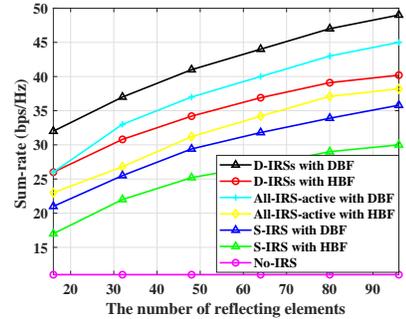}  
  \caption{$P=5~dBm$}
  \label{fig:(c)}
\end{subfigure}
\begin{subfigure}{.3\textwidth}
  \centering
  \includegraphics[scale=0.4]{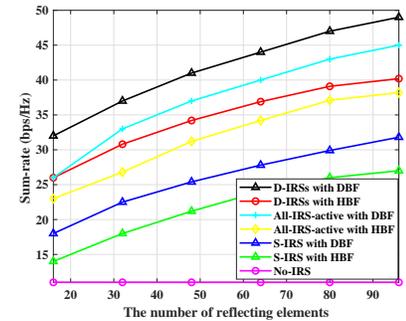}  
  \caption{$P=5~dBm$}
  \label{fig:(d)}
\end{subfigure}
\caption{Sum-rate versus the number of reflecting elements}
\label{fig:3}
\end{figure}

\begin{figure}
\centering
\begin{subfigure}{.2\textwidth}
  \centering
  \includegraphics[height=2.0in,width=1.4in]{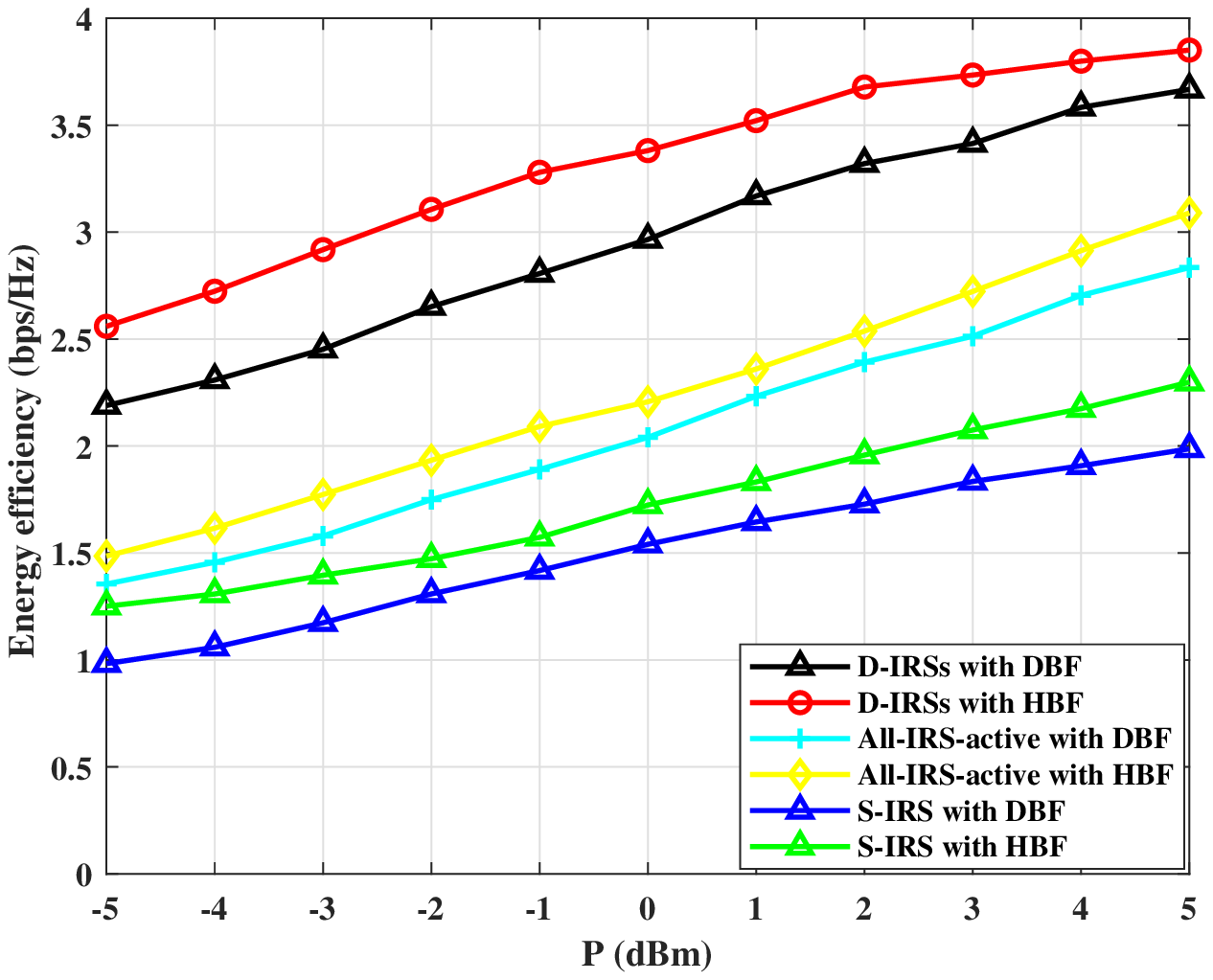}  
  \caption{$N_{r}=16$, $N_{RF}=8$}
  \label{fig:(a)}
\end{subfigure}
\begin{subfigure}{.2\textwidth}
  \centering
  \includegraphics[height=2.0in,width=1.4in]{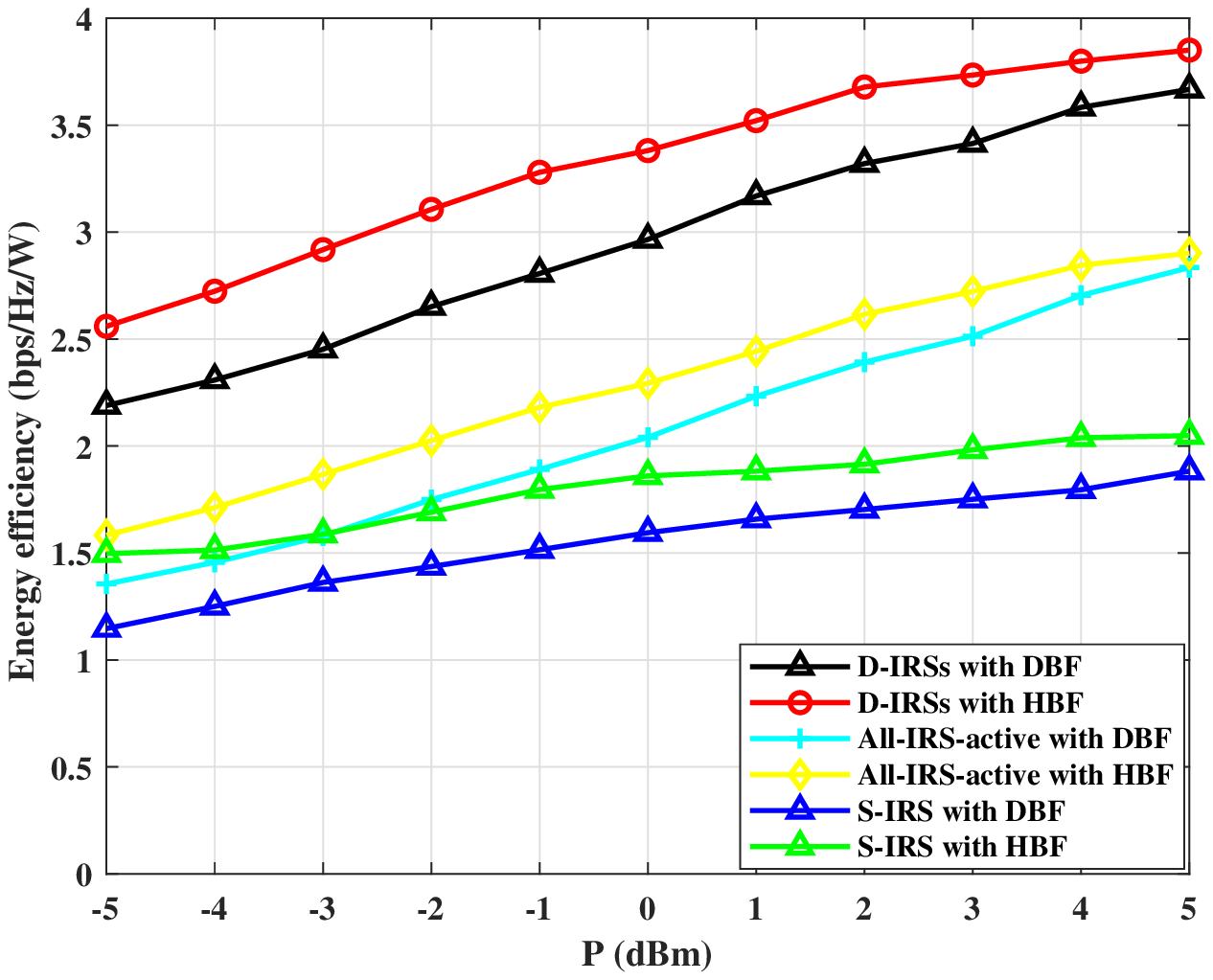}  
  \caption{$N_{r}=16$, $N_{RF}=8$}
  \label{fig:(b)}
\end{subfigure}
\caption{Energy efficiency versus transmit power $P$}
\label{fig:4}
\end{figure}
The IRS locates at $(0,60,20)$ and $(0,90,20)$ in the S-IRS scheme in Fig.~\ref{fig:3}(a) and Fig.~\ref{fig:3}(b), respectively. The sum-rate versus transmit power $P$ is shown in
Fig.~\ref{fig:3}(a) and Fig.~\ref{fig:3}(b). 
Based on the optimal beamformer obtained, it is easily proved that the sum-rate is an increasing function of $P$. In addition, as shown in Fig.~\ref{fig:3}(a) and Fig.~\ref{fig:3}(b), the sum-rate of the IRS-aided mmWave system with the HBF structure is lower than the IRS-aided mmWave system with digital beamforming (DBF). The reason is that the HBF obtained by using the OMP algorithm is an approximate solution, which leads to the degradation on the sum-rate.  
Thus, as $P$ increases, the sum-rate increases monotonously. The IRS locates at $(0,60,20)$ and $(0,90,20)$ in the S-IRS scheme in Fig.~\ref{fig:3a}(a) and Fig.~\ref{fig:3a}(b), respectively. Fig.~\ref{fig:3a}(a) and Fig.~\ref{fig:3a}(b) reveal that as reflecting elements increase from $16$ to $96$, the sum-rate increases monotonously. 
This is because the more reflecting elements result in sharper reflecting beams, thereby enhancing the sum-rate.  
Finally, from Fig.~\ref{fig:3a}(a) and Fig.~\ref{fig:3}(b), D-IRS scheme can increase up to 40\% sum-rate compared with the mmWave system with the S-IRS scheme, benefiting from the distributed deployment. Compared to the mmWave system with only one IRS, the D-RISs scheme can provide robust data-transmission since different RISs can be deployed geometrically apart from each other. Meanwhile, the D-RISs scheme can provide multiple paths with low path loss for received signals, which increases the received signal strength.



{}{Fig.~\ref{fig:4}(a) and Fig.~\ref{fig:4}(b) show the variation of the energy efficiency with the transmit power. The IRS locates at $(0,60,20)$ in the S-IRS scheme in Fig.~\ref{fig:4}(a), and the IRS locates at $(0,60,20)$ in the S-IRS scheme in Fig.~\ref{fig:4}(b). The simulation results show that with the increase of the BS transmit power, all the schemes increase with the increase of the transmit power and tend to be stable. This is because energy efficiency is not a monotonically increasing function of the transmit power, as shown in (\ref{24}). In addition, from the figure, we can find that the energy efficiency of the proposed D-IRSs scheme with HBF is larger than that of D-IRSs scheme with DBF, because the HBF requires fewer RF links than DBF, which will result in less RF power consumption. Compared with the All-IRS-active scheme and the S-IRS scheme, the proposed D-IRSs scheme have higher energy efficient, this is because the switch of D-IRSs can be adjusted adaptively the number of IRSs while increasing the sum-rate of the system, which can reduce the energy consumption of the IRS and increase the energy efficiency.}

\section{Conclusion}
In this paper, we have investigated the sum-rate maximization problem for a mmWave communication system with D-IRSs. 
The IRS phase shifts, the transmit beamforming at the BS, and IRS switch vector have been jointly optimized to maximize the sum-rate, subject to the constraints {}{on} the transmit power, minimum user rate, and unit-modulus constraints. 
To solve this problem, we have proposed the AO algorithm for the multi-users case. 
In particular, the joint design of the transmit beamforming and phase shifts was solved by using the SCA algorithm while the IRS switch vector was been obtained by using the greedy method. 
Finally, numerical results have shown that the proposed D-IRSs-aided scheme outperforms {}{the S-IRS scheme} in terms of the sum-rate and energy efficiency.

\ifCLASSOPTIONcaptionsoff
  \newpage
\fi


\bibliographystyle{IEEEtran}

\bibliography{relate}
\end{document}